\documentclass[showpacs,amsmath,amssymb,twocolumn,aps,showkeys,floatfix,prd,a4paper]{revtex4}
\usepackage{subfigure}
\usepackage{dcolumn}
\usepackage{bm}
\usepackage{epsfig}
\usepackage{amsfonts}
\usepackage{amssymb,amscd}
\def\lsim{\raise0.3ex\hbox{$<$\kern-0.75em\raise-1.1ex\hbox{$\sim$}}}
\def\gsim{\raise0.3ex\hbox{$>$\kern-0.75em\raise-1.1ex\hbox{$\sim$}}}

\newcommand{\be}{\begin{equation}}
\newcommand{\ee}{\end{equation}}
\newcommand{\ba}{\begin{eqnarray}}
\newcommand{\ea}{\end{eqnarray}}

\def\beq{\begin{equation}}
\def\eeq{\end{equation}}
\def\beqa{\begin{eqnarray}}
\def\eeqa{\end{eqnarray}}

\def\gappeq{\mathrel{\rlap {\raise.5ex\hbox{$>$}}
{\lower.5ex\hbox{$\sim$}}}}
\def\lappeq{\mathrel{\rlap{\raise.5ex\hbox{$<$}}
{\lower.5ex\hbox{$\sim$}}}}
\def\Toprel#1\over#2{\mathrel{\mathop{#2}\limits^{#1}}}

\begin{document}
\begin{flushright}
\vskip1cm
\end{flushright}

\title{Core destruction in knockout reactions }

\author{C.A. Bertulani}
\email{carlos.bertulani@tamuc.edu}
\affiliation{Institut f\"ur Kernphysik,  Technische Universit\"at Darmstadt, 64289 Darmstadt, Germany}
\affiliation{Department of Physics and Astronomy, Texas A\&M University-Commerce, 
Commerce, Texas 75429, USA
}

\begin{abstract}
A model is presented to calculate projectile core destruction in knockout reactions. It incorporates physics arguments similar to the formulation of the state of the art theory to calculate stripping and diffraction dissociation cross sections in heavy ion collisions with bombarding energies around 100 MeV/nucleon and larger. It is shown that secondary collisions between the incoming and struck nucleons with the projectile core decrease the core survival probability by as much as 9.5\%. However, no clear evidence is found for reduction of the cross section with increasing binding energy of the removed nucleon.   
\end{abstract}



\date{\today}

\maketitle

Direct reactions with radioactive nuclear beams with energies about 100 MeV/nucleon and larger have become an invaluable spectroscopic tool  in nuclear physics \cite{BERTULANI1993281}.  In particular, nucleon knockout reactions have been employed frequently to extract spectroscopic information on the evolution of single particle properties in nuclei as a function of their mass and charge  \cite{Hussein:1985,Bertulani:92,PhysRevLett.69.2050} (for a recent review, see \cite{AUMANN2021103847}). The analysis of nucleon removal, or knockout, reactions during the past 15 years has shown a systematic quenching of the extracted values of spectroscopic factors when compared to calculations based on the nuclear shell model \cite{gade2008,tostevin2014,AUMANN2021103847}.  This observation is often presented as an increasing reduction of the spectroscopic factors in terms of proton minus neutron separation energy, $\Delta S = \epsilon|S_n - S_p|$, where $\epsilon = +1$ for proton removal, and $\epsilon = -1$ for neutron removal. It remains a puzzle why this ``quenching'' has not been observed in the extraction of spectroscopic factors using transfer and (p,2p) experiments, which are also considered excellent probes of nuclear spectroscopy and very valuable for  studies of unstable nuclei using inverse kinematics  \cite{Flavigny:2012,Flavigny:2018,Lee:2010,lee2011,Atar:2018,Kaw18}.

Recent experimental works hinted at the possibility that state of the art theories aimed to describe nucleon removal reactions might miss consideration of important reaction channels such as the projectile core destruction in a secondary reaction of one outgoing nucleon stemming either from the projectile or from the target nucleus \cite{PANIN2019134802,Wamers2023}.  This possibility has been recently explored in Ref. \cite{gomezramos2023isospin} where, by invoking a non-local density formalism and compound nucleus formation and decay, the authors show that the effect of core destruction depends strongly on the binding energy of the removed nucleon, leading to a significant reduction of the cross section for deeply bound nucleons, and therefore reducing the isospin asymmetry of the spectroscopic factors observed in Refs. \cite{gade2008,tostevin2014,AUMANN2021103847}.  

In this work, we provide a simple theoretical mechanism to deal with core destruction, exploring the geometric nature of the reaction and including similar  physics arguments as those employed in past works in the derivation of theoretical expressions for knockout reactions   \cite{Hussein:1985,Bertulani:92,Bertulani:04,Bertulani:06}. The method includes minimal changes of the eikonal formalism adopted in past publications. Our numerical results imply that there is an evident reduction of the knockout cross sections due to core destruction by the participating nucleons. However, no clear dependence with the binding energy of the removed nucleon is found.

\begin{figure}[t]
\begin{center}
\includegraphics[
width=2.9in]
{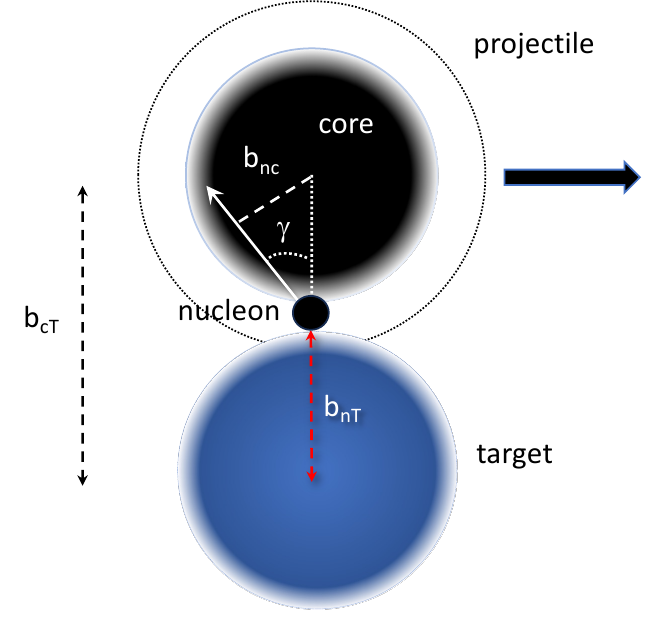}
\caption{Schematic drawing of the nucleon removal from a projectile followed by a rescattering with the projectile core.}
\label{schem}
\end{center}
\end{figure}

The widely employed expression to analyze experimental data of cross sections for single nucleon stripping from a projectile in a direct reaction is given by \cite{Hussein:1985,Bertulani:92,Bertulani:04,Bertulani:06,Hencken:1996}
\begin{eqnarray}
\sigma_{sp}^{str} &=& {1\over 2j+1} \sum_m\int \int d^2b_{nT} d^3r ( 1- |S_n(b_{nT})|^2) \nonumber \\
&\times&  \left\langle\psi_{jm}  \left|  | S_c(b_{cT})|^2 \right| \psi_{jm} \right\rangle, \label{strip}
\end{eqnarray}
where $S_n$ ($S_c$) is the S-matrix for the scattering of the nucleon (core) with the target at an impact parameter $b_{nT}$ ($b_{cT}$) (see Figure \ref{schem}). The nucleon wavefunction $\psi_{jm}({\bf r})$ depends on the single particle angular momentum of the removed nucleon and its projection,  $jm$, and on the nucleon internal coordinate written as ${\bf r}=(\boldsymbol{\rho},z)$. This coordinate is related to  $b_{nT}$ and $b_{cT}$ via $b_{cT}=|\boldsymbol{\rho} -{\bf b}_{nT}|$ \cite{Bertulani:04}. 
The total nucleon knockout cross section is given by 
\begin{equation}
\sigma_{kn}=\sum_j C^2S(j)f(A_P)\left[ \sigma_{sp}^{str}(j)+ \sigma_{sp}^{dd}(j) \right], \label{tod}
\end{equation}
where $C^2S(j)$ is the spectroscopic factor. The quantity $f$ is related to the mass number $A$ of the projectile by means of  $f(A) = A/(A-1)$ if $C^2S(j)$ is calculated with a many-body shell-model in a harmonic oscillator basis. In Eq. \eqref{tod}, $\sigma_{sp}^{dd}$ is the single-particle diffraction dissociation cross section given by
\begin{eqnarray}
\sigma_{sp}^{dd} = {1\over 2j+1} \sum_m\int d^2b_{nT}d^3r\Bigg[  \left\langle \psi_{jm}  \Big| \big| 1- S_nS_c  \big|^2 \Big| \psi_{jm}\right\rangle\nonumber \\
 -\sum_{m'}\Big| \left\langle \psi_{jm'}  \big| 1- S_nS_c \big| \psi_{jm'} \right\rangle\Big|^2 \Bigg].\label{diodes}
\end{eqnarray}
At high energies, the optical potential is connected to the nucleon-nucleon cross section and the expression for the S-matrix is \cite{Bertulani:04,Bertulani:06} 
\begin{equation}
S_i({\bf b_i}) =\exp \left[ -{\sigma_{NN}\over 4\pi} \int \rho_i(q) \rho_T(q) e^{-\beta_{NN}q^2} J_0(qb_i)q \,dq \right], 
\label{eik10}
\end{equation} 
where $J_{0}$ is the ordinary Bessel function of zeroth-order,  $\sigma_{NN}$ is the total nucleon-nucleon cross section, and $\beta_{NN}$ is the momentum dependence parameter. The nucleon density ($i=n$) is treated as a delta-function,  $\rho_n ({\bf r}) = \delta({\bf r})$, whereas  for $i= P$ and $i=c$ the ground state point nucleon densities of the projectile, $\rho_P({\bf r})$,  and the core, $\rho_c({\bf r})$, are used. $\rho_T({\bf r})$ is the point nucleon density distribution of the target. Eq. \eqref{eik10} uses the respective Fourier transform of these densities. The nucleon sizes are accounted for the nucleon-nucleon cross sections, $\sigma_{NN}$. 

Before continuing, it is worthwhile discussing the physics interpretation of the equations presented above. The stripping cross section has a very simple interpretation. It is an impact parameter integral of the product of the nucleon {\it removal} probability times the core {\it survival} probability. The diffraction dissociation cross section is derived by assuming that each cluster (nucleon or core) inside the projectile diffracts around the target. The combination of the two diffraction processes may impart a momentum transfer to the projectile, yielding distinct distorted waves for each cluster in the final state. When integrated over the momentum transfers and using completeness for the bound state $\psi_{jm} $ and the free distorted waves of the clusters one ends up with Eq.   \eqref{diodes} \cite{Hencken:1996}. For that matter, one needs to assume that the projectile has only one single bound state. This assumption is evidently not applicable to most nuclei studied in knockout reactions, and that is why the diffraction dissociation expression, Eq. \eqref{diodes}, widely utilized in the literature, relies on shaky grounds.

In the first instance, we consider that a nucleon from the projectile collides with a nucleon in the target. One of these nucleons has the possibility to induce a secondary collision with the core. The projectile loses one nucleon in the collision and the core might be destructed in the secondary collision. This could be only one of the various possible core destruction mechanisms, as suggested in Refs. \cite{louchart:2011:PRC,PANIN2019134802,Wamers2023}. If the core is destroyed, the observed nucleon removal cross section is smaller and the spectroscopic factors extracted from experiment will be larger.  The eikonal approximation allows for the treatment of secondary collisions in a straightforward  way and we adopt a formalism akin to that in  Ref. \cite{aumann2013} which has been very successful to analyze (p,2p) experiments.  A secondary collision of either one of the colliding nucleons with the core is expected to affect stripping much more than diffractive process. Diffractive dissociation is relatively soft and it unlikely that the nucleons will scatter sideways to the core. Hence, we concentrate on the stripping cross sections only.

The eikonal distortion up to a position ${\bf r} = ({\bf b},z)$ in coordinate space modifies the incoming wave $\Psi_{inc}$ as
\begin{eqnarray}
\Psi_{eik} ({\bf b},z) &=& G({\bf b},z) \Psi_{inc}(-\infty ),\nonumber \\
G({\bf b},z) &=& \exp\left[-{i\over \hbar v} \int_{-\infty}^z U_{opt}({\bf b},z')dz'\right] , \label{green}
\end{eqnarray}
where $v$ is the projectile velocity, and $U_{opt}$ is the optical potential. If the above integral is extended to infinity, one recovers the S-matrices used in Eqs. \eqref{strip} and \eqref{diodes} as $S_i({\bf b})= G_i({\bf b},\infty)$. Eq. \eqref{green} shows that in the eikonal approximation, the scattering wave acquires a phase up to position ${\bf r}=({\bf b},z$). This can readily be used to propagate sequential particle scattering to treat multiple collisions by recursively calling different points ${\bf r}$ as a new source of subsequent eikonal waves, i.e.,
 \begin{eqnarray}
G ({\bf b},z) &=& \int^z dz' dz''   d{\bf b}'d{\bf b}''  \cdots G({\bf r}') G({\bf r}'')\cdots   \label{green2}
\end{eqnarray}
Although this formalism is attractive and has been reported in the literature to treat multiple collisions using eikonal waves \cite{PhysRev.103.443,PhysRev.184.1745}, it is important to notice that  it does not apply to the case of core destruction studied here because the two steps are independent. 

\begin{figure}[t]
\begin{center}
\includegraphics[
width=2.9in]
{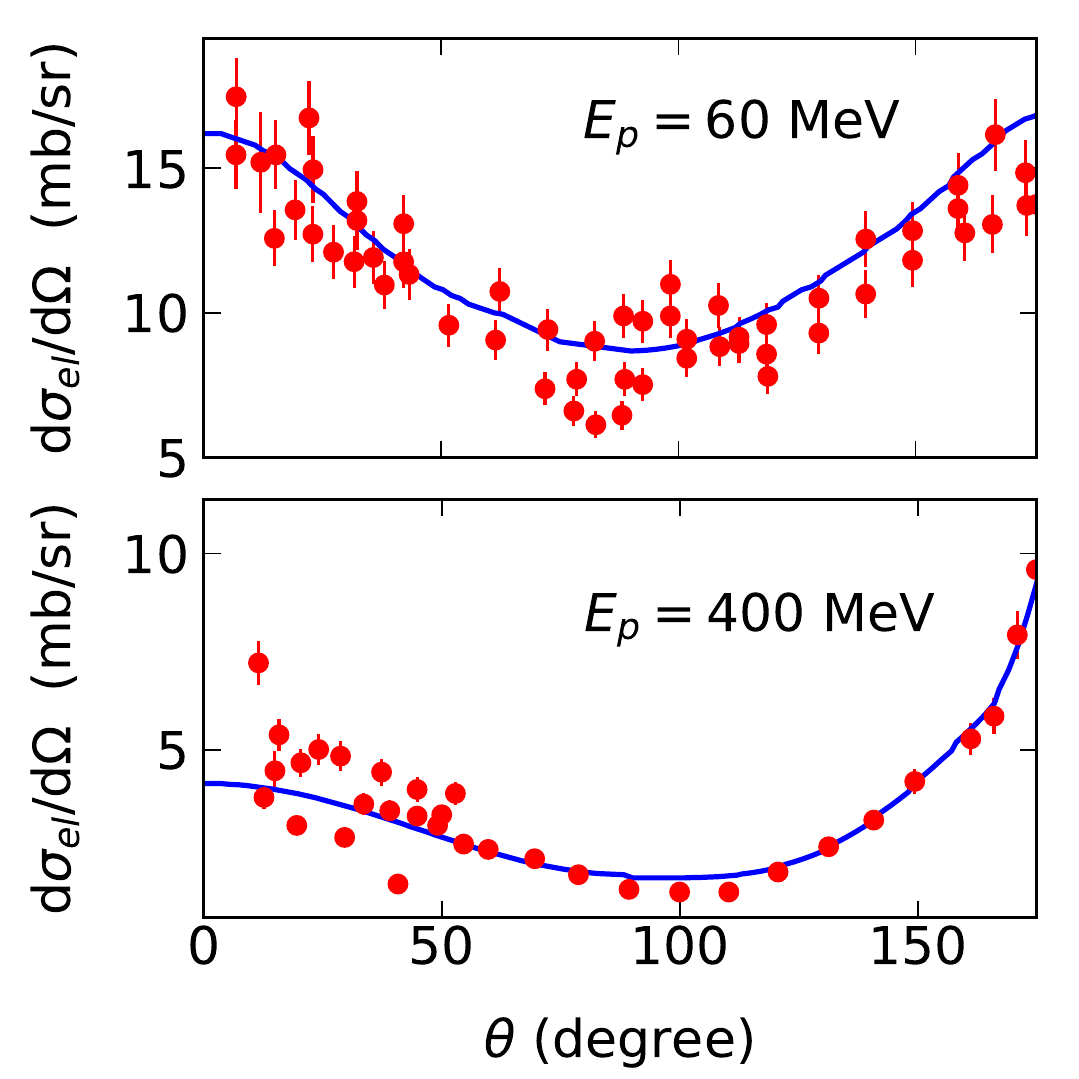}
\caption{Differential cross sections for proton-proton collisions as a function of the scattering angle in the center of mass. Data is retrieved from EXFOR library \cite{OTUKA2014272} and curves are parametrizations taken from  Refs.  \cite{CUGNON1996215,CUGNON1997475}.} 
\label{nnscat}
\end{center}
\end{figure}

Instead of the multistep formalism delineated above, we use a much simpler procedure based on the same arguments leading to Eq. \eqref{strip} to derive the stripping cross section including a correction due to core destruction by a secondary scattering. We  propose modifying the integrand of Eq. \eqref{strip} to  
\begin{eqnarray}
( 1- |S_n|^2)| S_c|^2 \rightarrow ( 1- |S_n|^2)\left| S_c\right|^2\left(1 - \left<|S_{nc}|^2\right>\right)  , \label{sccorr}
\end{eqnarray}
with
\begin{equation}
 \left<|S_{nc}|^2\right> = {1\over \sigma_{NN}^{el}} \int d\Omega {d\sigma_{NN}^{el}(\theta)\over d\Omega}
 |S_{nc}(b_{nc}(\theta,\phi))|^2 ,\label{intrs}
 \end{equation}
 where $\left< ... \right>$ denotes an average over scattering angles. The S-matrix in Eq. \eqref{intrs} is calculated in the same way as in Eq. \eqref{strip} (i.e., using Eq. 4), but with the impact parameter replaced by $b_{nc}$. The main assumption is that the secondary collision yields a nucleon with the same center of mass energy as the incoming nucleon. The secondary nucleon which stems either from the projectile or the target due to the binary nucleon-nucleon collision, enters the core with an angle $\gamma$, as schematically shown in Fig. \ref{schem}.   The impact parameter $b_{nc}$ entering Eq.  (8) is now dependent on the scattering angle $\gamma(\theta,\phi)$, where the dependence of $\gamma$ on $\theta$ and $\phi$ is defined below.
The additional term,  $1-\left<|S_{nc}|^2\right>$, is the probability that one of the nucleons collides with the core which is likely to destroy it, depending on the impact parameter.  The integration in Eq. \eqref{intrs} runs over all angles, or equivalently, all impact parameters $b_{nc} (\theta, \phi)$ that lead to a nucleon scattering toward the core. The first nucleon-nucleon collision is assumed to be quasi-free, i.e., an elastic collision, thus preserving the center-of-mass energy. From this point on, the secondary nucleon, either from the projectile or from the target, is again described by an eikonal S-matrix when it scatters with the core. These are similar assumptions as those used in Ref. \cite{{aumann2013}} to describe secondary collisions in (p,2p) reactions.

\begin{figure}[t]
\begin{center}
\includegraphics[
width=3.4in]
{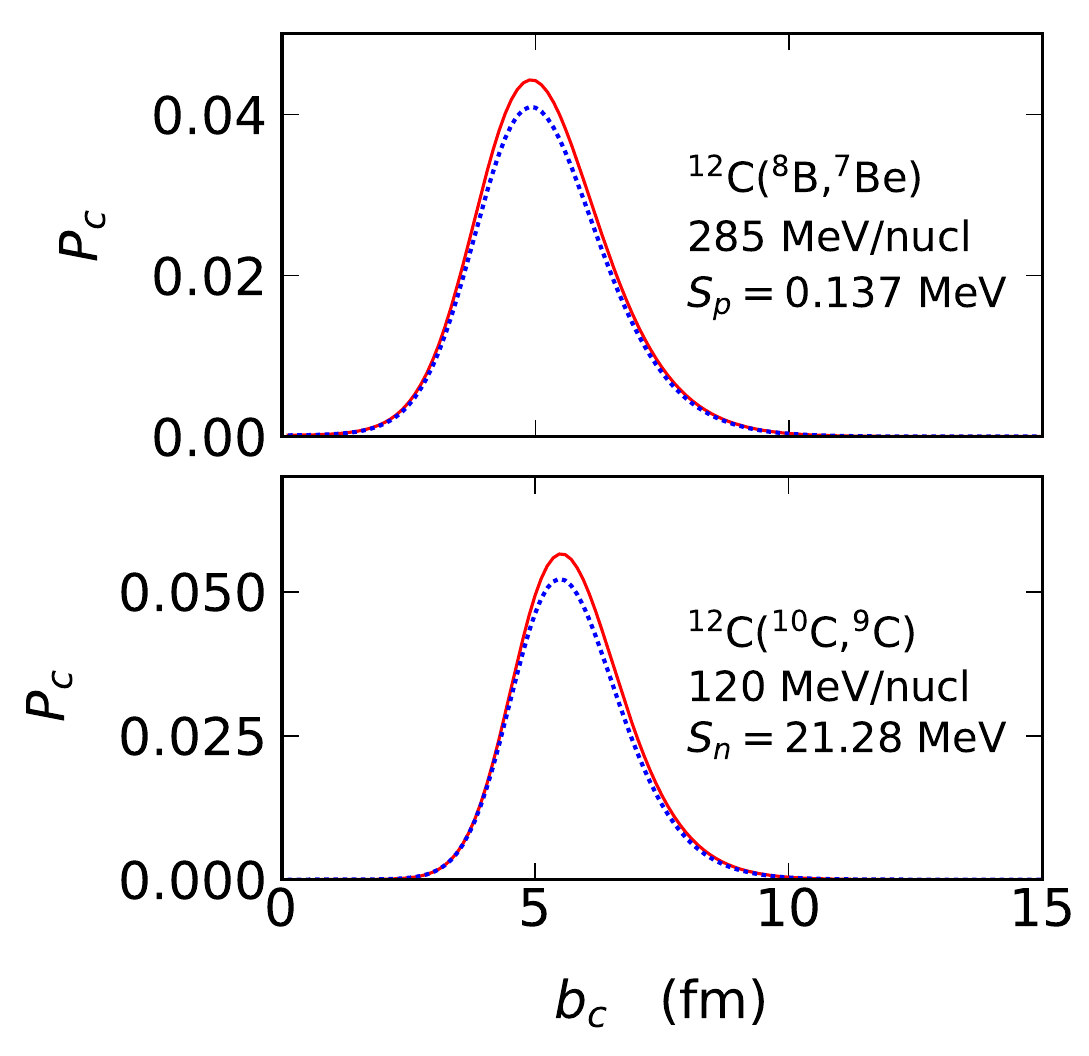}
\caption{Core survival probabilities, $P_c(b_c)$, defined according to Eq. \eqref{strip2} (solid curves) and with Eq. \eqref{strip3} (dotted curves).  Upper panel: $^{12}$C($^8$B,$^7$Be) at 285 MeV/nucleon with a proton separation energy $S_p=0.137$ MeV, angular momentum $J^\pi = 3/2^-$. Lower panel: $^{12}$C($^{10}$C,$^9$C) at 120 MeV/nucleon with a neutron separation energy $S_n=21.28$ MeV, angular momentum $J^\pi = 3/2^-$. The dotted (solid) lines include (do not include) core destruction due to a secondary collision of one of the colliding nucleons and the core.}
\label{prob}
\end{center}
\end{figure}

For simplicity, we assume that the most likely position where the projectile nucleon is struck occurs when $z=0$; the closest distance the projectile passes by the target.  Figure \ref{schem} shows a schematic drawing of the nucleon-core rescattering process. Although medium effects on nucleon-nucleon collisions might be important, we determine the scattering angle using the free- nucleon-nucleon elastic cross section $ d\sigma_{NN}^{elast}/ d\Omega$ as a weight function in the angular average. Evidently, only a limited range of angles $\gamma(\theta,\phi)$ leads to the core destruction.  The scattering angle $\gamma$ can be related to $b_{nc}$ and the incoming nucleon and core impact parameters through $b_{nc}=|{\bf b}_{cT}-{\bf b}_{nT}| \tan\gamma$, where $\gamma$ is defined as the angle between the nucleon-nucleon scattering direction and the x-axis, perpendicular to the z-axis beam direction. The angle $\gamma$ and the nucleon-nucleon scattering angles $(\theta, \phi)$ are related by $\cos\gamma=\sin\theta \cos\phi$, where $\phi$ is the off-plane azimuthal angle.  As indicated in Fig. \ref{schem}, the scattered nucleons will hit the core in a direction nearly perpendicular to the beam direction. 

The experimental pp and np differential elastic scattering cross sections were obtained from the large amount of experimental data that can be accessed at the EXFOR library \cite{OTUKA2014272}. For all energies considered in this work, we use a parametrization of the differential cross sections found in Refs.  \cite{CUGNON1996215,CUGNON1997475} which fits the experimental data quite well.   As an example, in Fig. \ref{nnscat} we show the differential cross section for proton-proton scattering at beam energies of 60 and 400 MeV/nucleon and as a function of the scattering angle in the center of mass. It is clear that the cross sections peak at forward and backward angles, but there is a non-negligible probability that the protons scatter around 90$^0$, thus hitting the projectile core.  We also carry out an isospin average of the differential cross section according to $\left< \sigma_{pN}^{el}(\theta) \right> = (Z_T \sigma_{pp}^{el}(\theta) +  N_T \sigma_{pn}^{el}(\theta) )/A_T$ for proton knockout, with an equivalent expression for neutron knockout, exchanging $N_T$ and $Z_T$.   A similar average is done for the total scattering cross section appearing in Eq. \eqref{intrs} which accounts for nuclear absorption.

In Fig. \ref{prob} we plot the core survival probabilities, $P_c(b_c)$, for two reactions with very different nucleon separation energies. $P_c$ is defined by
\begin{equation}
P_c(b_{cT}) = \int  d^3r ( 1- |S_n(b_{nT})|^2)  \left\langle\psi_{jm}  \left|  | S_c(b_{cT})|^2 \right| \psi_{jm} \right\rangle, \label{strip2}
\end{equation}
without rescattering correction, or
\begin{eqnarray}
P_c(b_{cT})  &=& \int  d^3r ( 1- |S_n(b_{nT})|^2)  \left(1 - \left<|S_{nc}|^2\right>\right) \nonumber \\ &\times&\left\langle\psi_{jm}  \left|  | S_c(b_{cT})|^2 \right| \psi_{jm} \right\rangle, \label{strip3}
\end{eqnarray}
when core-destruction is included. $b_{nT}$ is expressed in terms of $b_{cT}$ and ${\bf r}$, as explained in the text following Eq. \eqref{strip}.
 
The core survival probability, after a nucleon removal from the projectile, is shown in Fig. \ref{prob}  (upper panel) for $^{12}$C($^8$B,$^7$Be) at 285 MeV/nucleon with a proton separation energy $S_p=0.137$ MeV, angular momentum $J^\pi = 3/2^-$. In the lower panel we consider the reaction $^{12}$C($^{10}$C,$^9$C) at 120 MeV/nucleon with a neutron separation energy $S_n=21.28$ MeV, angular momentum $J^\pi = 3/2^-$. The dotted (solid) lines include (do not include) core destruction due to a secondary nucleon collision with the core. 

\begin{figure}[t]
\begin{center}
\includegraphics[
width=3.4in]
{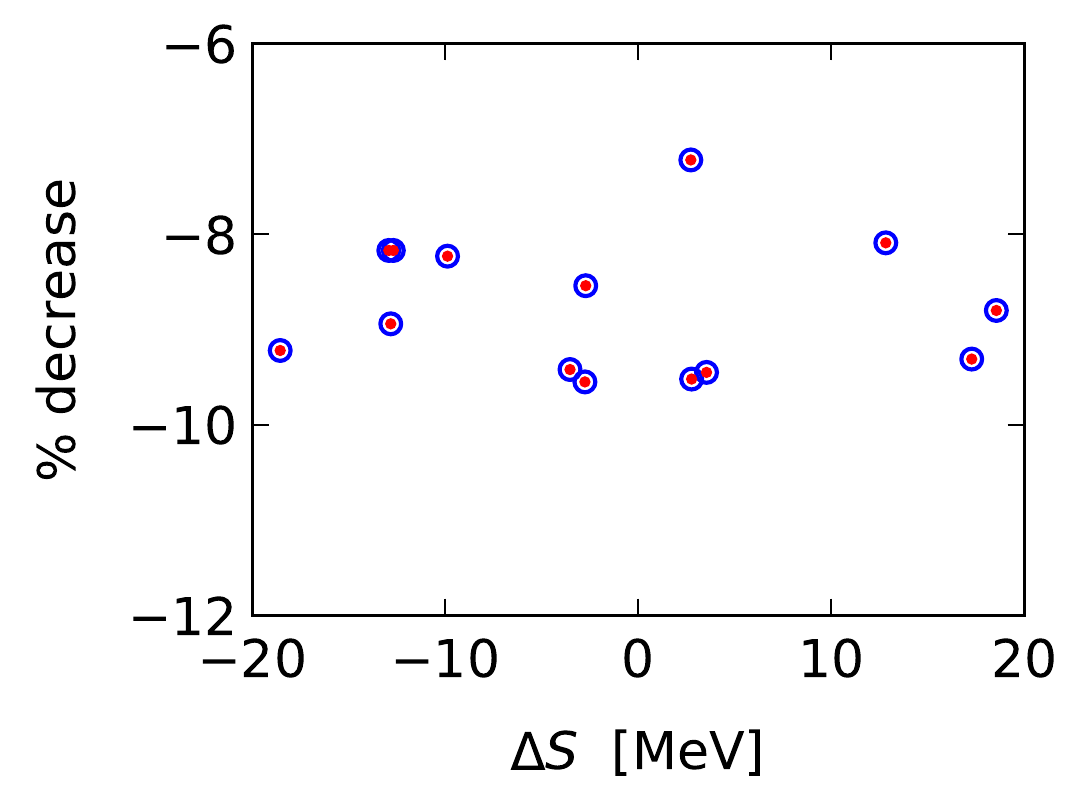}
\caption{Decrease of stripping cross sections as a function of $\Delta S=S_p-S_n$ for neutron removal or $\Delta S=S_p-S_n$ for proton removal.}
\label{quench}
\end{center}
\end{figure}

An important remark is that our calculations are based on ab initio many-body wavefunctions  using the No-Core Shell Model (NCSM) formalism, as described in Ref. \cite{PhysRevC.105.024613}. It has been recently found out that the spectroscopic factors extracted from  (p,2p) as well as from knockout reactions are more dependent on the interior part of the single-particle wavefunction than initially expected \cite{PhysRevC.104.L061602,PhysRevC.105.024613}. Fig. \ref{prob}  shows that there is  a non-negligible correction of the core survival probabilities as a function of the core impact parameter in collisions with $^{12}$C targets. The correction is largest at the maximum of the probability which occurs at core impact parameters $b_c \sim R_c + R_T$, where $R_c$ ($R_T$) is the mean-square radius of the core (target).

In Table \ref{SF10} we show the calculated values for one-nucleon-removal ({\it stripping}) cross sections for selected single-particle states, most of them already examined in Ref. \cite{PhysRevC.105.024613}. For clarity, we recall that $\Delta S=S_p-S_n$ for neutron removal or $\Delta S=S_p-S_n$ for proton removal. The separation energies are not the experimental ones, but those obtained with the NCSM calculations for the respective states \cite{PhysRevC.105.024613}. It is evident that the correction due to  secondary scattering does not display a visible correlation with the separation energies, as is also seen in Fig. \ref{quench}. The correction due to core destruction is  reflected in the total nucleon knockout cross sections, all decreasing by nearly the same percentage.

\begin{widetext}
\begin{center}
\begin{table}[!htb]
\centering
\caption{Theoretical values for one-nucleon-removal stripping cross sections for selected single-particle states. $\Delta S=S_p-S_n$ for neutron removal or $\Delta S=S_p-S_n$ for proton removal.}
\setlength{\tabcolsep}{1.8mm}{
\begin{tabular}{lccccccc} \hline \hline
 Reaction           & $E_{beam}$ & $S_p[S_n]$&$\Delta  S$ &  $J^\pi(j)$ &$\sigma_{str}$ & $\sigma_{str}^{rescatt}$ &$\%$ change \\  \hline
                   & MeV/nucleon &MeV &MeV &    mb & mb &  \\ \hline
$^{9}$B($^7$Li,$^6$He)& 80  & 9.98& 2.723  & $0^+(3/2)$&  20.49 & 19.01 & -7.22  \\
$^{9}$B($^7$Li,$^6$Li)  &120 & 7.25 & -2.723 & $1^+(3/2)$&  24.23 & 22.16 &  -8.54  \\
$^{12}$C($^8$B,$^7$Be)  &285 & 0.137& -12.690  &   $3/2^-(3/2)$&  42.49& 39.02 & -8.17 \\
$^{12}$C($^9$C,$^8$B)  & 78   &1.3& -12.925  &   $2^+(3/2)$& 40.14 &36.86 & -8.17\\
$^{12}$C($^9$Li,$^8$Li)  &100   &  4.06&-9.882  &  $2^+(3/2)$& 40.32 & 37.00 &  -8.23 \\
$^{9}$Be($^{10}$Be,$^9$Li)  &  80 &19.64  &12.824  &  $3/2^-(3/2)$& 35.33 & 32.47 & -8.09  \\
$^{9}$Be($^{10}$Be,$^9$Be)  &  120& 6.812&-12.824  &   $3/2^-(3/2)$&  77.62 & 70.68 & -8.94  \\
$^{9}$Be($^{10}$C,$^9$C)  &  120   & 21.28&17.277  &  $3/2^-(3/2)$&44.57 & 40.50 & -9.31  \\
$^{12}$C($^{12}$C,$^{11}$B)  &250   &15.95& -2.764  &   $3/2^-(3/2)$&64.68 & 58.70& -9.55  \\
$^{12}$C($^{12}$C,$^{11}$C)  & 250   & 18.72& 2.764  &  $3/2^-(3/2)$& 74.16 & 67.10 & -9.52 \\
$^{12}$C($^{14}$O,$^{13}$N)  &  305   &1.531& -18.552  &   $1/2^-(1/2)$&37.45 & 33.99 & -9.22  \\                            
$^{9}$Be($^{14}$O,$^{13}$O)  &    53   &3.234& 18.552  & $3/2^-(3/2)$&   25.57 & 23.32 & -8.80  \\
$^{12}$C($^{16}$O,$^{15}$N)  &   2100 &22.04&-3.537  &   $3/2^-(3/2)$&  46.90 & 42.48 &-9.42  \\
$^{12}$C($^{16}$O,$^{15}$O)  &  2100   & 22.04&  3.537  &  $3/2^-(3/2)$& 44.46 & 40.26 & -9.45\\
\hline \hline
\end{tabular}}\label{SF10}
\end{table}
\end{center}
\end{widetext}

We conclude that rescattering of either the incoming or the struck nucleon by the core tends to decrease the total nucleon knockout cross sections by as much as 9.5\%, at least for the selected cases chosen in this work. This is an important result as it implies that the spectroscopic factors extracted from knockout reactions need revisions. They are probably larger than they should be because the spectroscopic factors have been extracted and reported in the literature using primarily the Eqs. (\ref{strip}-\ref{eik10}). However, we do not reproduce the energy dependence with the binding nucleon energy, which would pave the way to solve the so-called ``quenching puzzle'', as suggested in Ref. \cite{gomezramos2023isospin},  because in our model rescattering effects tend to reduce the stripping cross sections without any correlation with $\Delta S$. Our model is consistent with the physical arguments employed in the derivation of Eqs. (\ref{strip}-\ref{eik10}), where we have included an additional term to account for core destruction. The model relies on a geometrical condition which is independent of the binding energy of the removed nucleon. Perhaps, the most important effect leading to a possible solution of the ``quenching puzzle''  is the proper use of single particle wavefunctions obtained with ab initio many-body models, as claimed before in Refs. \cite{PhysRevC.104.L061602,PhysRevC.105.024613}.  As shown in those references, the quenching  of the spectroscopic factors tends to weaken when Eq. \eqref{strip} uses proper ab initio wavefunctions. The effect of the $S_n$ and $S_c$ eikonal matrices in the integrand of Eq. \eqref{strip} is to ``chop off''  parts of the bound-state wavefunction. It has been shown in past publications that this leads to  a binding energy dependence of the cross sections. But Refs. \cite{PhysRevC.104.L061602,PhysRevC.105.024613}  show there is also a non-negligible dependence of the knockout cross sections on the internal part of the single-particle wavefunction $\psi_{jm}$, and this weakens the quenching of spectroscopic factors. The introduction of the core-destruction effect considered here is a geometric absorption of a scattered nucleon with the core that does not depend on the nucleon binding energy. In fact, the secondary scattering proceeds only after the valence nucleon is removed from  the projectile. Therefore, no additional energy dependence is expected to occur.

\begin{acknowledgments}

The author has benefitted from useful discussions with Leonid Chulkov, Bj\"orn Jonson, Thomas Aumann and Alexandre Obertelli. He acknowledges support by the U.S. DOE grant DE- FG02-08ER41533 and the Helmholtz Research Academy Hesse for FAIR.

\end{acknowledgments}

\hspace{1.0cm}

\end{document}